# Non-Linear Photocurrent Response to Bosonic Final State Stimulation in Microcavity Diodes

Aniruddha Bhattacharya[1, *], Evgeny Sedov[2, 3, †], Md Zunaid Baten[1], Pallab Bhattacharya[1, §], and Alexey Kavokin[2, 4, 5, ‡]

[1] *The Solid-State Electronics Laboratory,*

*Department of Electrical Engineering and Computer Science, University of Michigan,*

*1301 Beal Avenue, Ann Arbor, Michigan 48109, USA*

[2] *School of Physics and Astronomy, University of Southampton, SO17 1NJ Southampton, United Kingdom*

[3] *Vladimir State University named after A. G. and N. G. Stoletovs, Gorky str. 87, 600000, Vladimir, Russia*

[4] *CNR-SPIN, Viale del Politecnico 1, I-00133, Rome, Italy*

[5] *Spin Optics Laboratory, St. Petersburg State University, Ul'anovskaya 1, Peterhof, St. Petersburg 198504, Russia*

[*] anirudb@umich.edu
[†] evgeny_sedov@mail.ru
[§] pkb@umich.edu
[‡] A.Kavokin@soton.ac.uk

**ABSTRACT**

We report the optical excitation-dependent output photocurrent characteristics of GaN-based polariton diode lasers operated under reverse-bias at room temperature. The photocurrent demonstrates a non-linear enhancement at an incident optical power of ~ 1.6 mW, which is approximately equivalent to the value of polariton lasing threshold observed when the diodes are operated under forward bias conditions. This is explained in the framework of an Auger-like process of excitonic dissociation into its constituent electron-hole pairs, which can be stimulated by the occupation of the polariton lasing states. The observed effect is a remarkable manifestation of the bosonic final state stimulation in polariton lasers. A model based on the coupled kinetic equations for the free carriers, the excitonic reservoir and the polariton condensate shows a good agreement to the experimental data.

A polariton laser is one of the most important outcomes of the research field of Polaritonics: physics of strongly coupled light-matter systems [1]. In this light source, coherent emission is produced by the spontaneous radiative recombination from a macroscopic, coherent, degenerate and non-equilibrium exciton-polariton condensate [2, 3]. The non-linear threshold of a polariton laser is lower than that of conventional semiconductor photon lasers usually by one to three orders of magnitude [4-8], mainly because it does not require the inversion of electronic population. Photocurrent measurements on polariton devices have been previously reported [9-16]. Photocurrent spectra recorded from a reverse-biased microcavity diode can directly map the polariton dispersion characteristics [10, 13]. The optical nonlinearity in a similar device due to switching between strong and weak coupling has been reported by Bajoni et al. [11]. The non-linearity in the reverse biased photocurrent due to a blue shift of the lower-polariton (LP) branch into resonance with the optical pump energy and consequent increase in injection efficiency has been reported by Winkler et al. [14]. A change in the slope of the lateral photocurrent, measured in the in-plane direction, was reported by Brodbeck et al. and attributed to a transition from bosonic (polariton) to fermionic (photon) lasing [15]. All these measurements were performed at cryogenic temperatures. In this Letter, we report, for the first time, the observation of a non-linear dependence of the photocurrent on pumping intensity in a microcavity diode at room temperature, as a new manifestation of bosonic final state stimulation in polariton lasers [17-23]. The excitation density measured at the onset of the non-linearity coincides with the threshold current density for polariton lasing in the same device. This observation also serves as a proof of principle of an optically-controlled low-energy switching device, where the output photocurrent of a reverse-biased GaN-based microcavity diode is controllably switched between a low output current value (or, nominally OFF state) and high output current value (or, nominally ON state), by pulsed optical excitation at normal incidence. A robust, ultra-fast and energy-efficient all-optical, all-electrical or an electro-optical switching device can potentially have a wide range of practical applications in information and signal processing including Application Specific Integrated Circuits (ASICs) and Systems-on-a-Chip (SOCs), on-chip communications and signal routing, control and sensing. In the present study, we have investigated the excitation-dependent output photocurrent characteristics of bulk GaN-based microcavity polariton diode lasers [8, 24-26]. The observed non-linear dependence of the photocurrent on the excitation intensity is reproduced by theoretical calculations including an Auger-like process of excitonic dissociation into its constituent electron-hole pairs, which can be stimulated by the occupation of the polariton lasing states.

The present devices are identical to the polariton diode lasers characterized and reported by us previously [25, 26], wherein the device processing and the measured characteristics of the lasers have been described in detail. All of these devices originate from the same epitaxial sample and have been processed identically and simultaneously. The defect density in the active region of the devices is $\sim 6.1 \times 10^8$ cm$^{-2}$ [25]. A schematic representation of the microcavity diode along with the device heterostructure is shown in Fig. 1(a). As has been highlighted in our previous publications [8, 24-26], these devices employ a novel lateral injection architecture, wherein the directions of current flow and polariton emission are mutually orthogonal. This enables the realization of a relatively low series resistance microcavity diode along with a reasonably good quality factor cavity [8]. This is particularly important in the context of the experiments reported in the present Letter, where the effects of polariton lasing on the diode current have been studied. To start with, let us recall some of the previously reported polariton lasing characteristics. The devices are characterized by a cavity photon-to-exciton detuning δ ranging from − 4 to − 13 meV, and a strong coupling in the microcavity characterized by a Rabi splitting Ω of ~ 33.9 to ~ 35.5 meV. The microcavity quality factor Q and the cavity mode lifetime $\tau_C$ were found to vary from ~ 1700 to ~ 2100, and from ~ 0.3 ps to ~ 0.4 ps, respectively. The polariton laser thresholds varied from ~ 125 A/cm$^2$ to ~ 375 A/cm$^2$ [26]. In addition, the onset of non-linear emission was accompanied with the collapse of the LP emission linewidth and a small blueshift of the peak electroluminescence energy. The latter, which is primarily caused by the repulsive self-interaction of the lower-polaritons, is particularly important as it signifies the particulate nature of the radiative species. In contrast, photon lasers do not show any blueshift in their peak emission energy, unless they are injected at very high densities where band-filling effects become important. The redistribution of the lower-polaritons in momentum space as a function of the injection current was also studied. A random and non-thermal LP occupation below threshold was found to transform into a peaked occupancy at $k_{\parallel} \sim 0$ above threshold. In addition, there was no evidence of a polariton relaxation bottleneck at any injection. At threshold, the occupancy was analyzed by a Maxwell-Boltzmann distribution to yield the effective polariton temperature of $T_{LP} \sim 270$ K [25]. It was also observed that the emission is unpolarized below the nonlinear threshold and is linearly polarized above it with a maximum polarization of ~ 22%. Conventional cavity mode-mediated photon lasing was also observed in one of the polariton laser devices at an injected current density of ~ 36.8 kA/cm$^2$ [26]. All measurements had been made at room temperature. Some of these results are described in the Supplemental Material [27] for completeness. In particular, the strong coupling characteristics are shown for a similar device having a similar value for the ground-state LP mode energy of ~ 3.37 eV, as those considered in the present study.

Measurements in the present study have been done on two structurally identical microcavity diodes, which we have labeled as Device 1 and 2. The light (output)-current (L-I) characteristics of Device 2, in the normal direction ($k_{\parallel} \sim 0$), are determined by recording the output

LP electroluminescence intensities by a photomultiplier tube after spectral filtering through an imaging monochromator, as a function of continuous wave injection current density are shown in Fig. 1 (b). The ground state occupation numbers, quoted in Fig. 1(b), are estimated by normalizing the recorded electroluminescence intensities to unity at the threshold. The peak emission wavelength is $\lambda \sim 367$ nm (see Supplemental Material [27]). The polariton lasing threshold current densities of Device 1 and 2 are $\sim 1993$ A/cm$^2$ and $\sim 2899$ A/cm$^2$, respectively. The sub-threshold slopes of these characteristics, when plotted in a double logarithmic scale, are $\sim 0.45$ and $\sim 0.48$ respectively. These rather small values are indicative of device degradation. Nevertheless, the non-linear slopes above the threshold are $\sim 2.95$ and $\sim 3.68$, respectively, which are in reasonably good agreement with the values we have previously observed and reported, for the slopes in the polariton lasing regimes. Device 2 shows slightly stronger non-linearity as compared to Device 1. Thus, Device 2 has been described in the main text, whereas the output characteristics of Device 1 are shown in the Supplemental Material [27]. The observation of a larger polariton laser threshold as compared to what we have previously reported in identical GaN-based devices ($\sim 125$ A/cm$^2$ to $\sim 375$ A/cm$^2$ [26]) is also indicative of deterioration in the electrical injection process. The increase in the thresholds is, in all probability, a complex interplay of increased density of non-radiative recombination and trapping centers, increased carrier leakage from the active region, and reduction in the carrier injection efficiency at the contacts. This has been taken into account in our theoretical calculations. Nevertheless, as discussed later, the injected carrier density corresponding to the threshold in both of the devices is below the excitonic Mott transition as well as the transparency carrier density in the medium.

Optical excitation-dependent photocurrent measurements are performed at room temperature on the polariton lasers under a very small reverse bias (see Supplemental Material [27]). The experimental configuration is shown in Fig. 1(a). Measurements have not been done under forward bias, as the diode diffusion current would overwhelm the relatively weaker photocurrent response. The excitation-dependent output photocurrent characteristics of Device 2 are shown in Figs. 2 (a) and (b) for different values of the applied reverse bias. We observe a non-linear enhancement of the absolute magnitude of the photocurrent at incident optical pump powers $I_p$ ~ 1.6 mW. Similar behavior has also been observed for Device 1 (see Supplemental Material [27]). This effect is most clearly manifested at values of reverse biases of ~ 0.1 and 0.2 V. At higher values, the non-linearity starts getting obscured by the increased magnitude of the dark current of the diode. The photocurrents plotted in Figs. 2(a) and (b), are the as-measured values. We would like to emphasize that these currents are the true reverse currents flowing through the diodes, originating from photo-generated carriers which diffuse to the depletion region, wherein they are accelerated and are finally collected at the ohmic contacts of the device. Thus, our measurements are different from those reported by Brodbeck et al. [15], where a lateral photocurrent, proportional to the in-plane quantum-well carrier density, is collected by Schottky contacts formed across the device active region. The pump powers at the onset of photocurrent non-linearity correspond to a LP density $N_{3D}$ ~ 3 X $10^{16}$ cm$^{-3}$ , estimated with the relationship: $N_{3D}$ = $(1-R_{NI})A_{NI}I_p/(f_p\hbar\omega\pi r^2 L)$ [28] where $R_{NI}$ is the cavity reflectance at normal-incidence of the strongly coupled microcavity estimated to be ~ 0.9, $A_{NI}$ is the overall conversion efficiency of the incident photons to intra-cavity polaritons at normal incidence estimated to be ~ 0.46 as discussed later, $I_p$ is the incident pump power, $f_p$ is the incident optical excitation repetition rate of the mode-locked laser ~ 80 MHz, $\hbar\omega$ is the energy of the pump photons ~ 3.378 eV at the LP ground-state energy, $r$ is the radius of the pump spot size ~ 5 µm and $L$ is the device cavity length ~ 690 nm.

This value of $N_{LP}$ is lower than the transparency density of $\sim 10^{18}$ cm$^{-3}$ [29] and the exciton Mott density of $\sim 10^{19}$ cm$^{-3}$ in GaN [30] by two and three orders of magnitude, respectively.

We believe that the observed non-linearity in the photocurrent is related to the onset of stimulated scattering and polariton lasing triggered by the optical excitation. To model the kinetics of the system for analyzing the experimental data in the most general terms, we use a set of rate equations for the occupancies of the coherent polariton state $N_C(t)$, the reservoirs of active excitons, $N_X(t)$, and free carriers, $N_{eh}(t)$:

$$\frac{dN_C}{dt} = (RN_X + AN_X^2)(N_C + 1) - g_C N_C, \quad (1)$$

$$\frac{dN_X}{dt} = W_O N_{Oeh} + W_J N_{Jeh} - [g_X + (R + 2AN_X)(N_C + 1)]N_X, \quad (2)$$

$$\frac{dN_{Oeh}}{dt} = P_O + AN_X^2(N_C + 1) - (g_{eh} + W_O)N_{Oeh}, \quad (3)$$

$$\frac{dN_{Jeh}}{dt} = P_J - (g_{eh} + W_J)N_{Jeh}. \quad (4)$$

In Eq. (1), $R$ is the acoustic phonon assisted relaxation rate from the reservoir to the condensate. $A$ is the Auger scattering constant describing the Auger-like process of the scattering of the exciton to the condensate accompanied by the dissociation of one of the scattered excitons. $\gamma_C$, $\gamma_X$ and $\gamma_{eh}$ are the relaxation rates of the condensate polaritons, reservoir excitons and free carriers, respectively. Here we consider the reservoir of free carriers consisting of two parts, $N_{eh} = N_{Oeh} + N_{Jeh}$, where $N_{Oeh}$ and $N_{Jeh}$ describe densities of the reservoirs created by optical and electrical pumping with the pump powers $P_O$ and $P_J$, respectively. Having a different origin, the reservoirs of free carriers pump the reservoir of excitons $N_X$ with different rates, $W_O$ and $W_J$, respectively;

$W_O > W_J$. The optically pumped reservoir is also replenished by the Auger-like process. Here we leave the spatial degree of freedom beyond the scope of consideration assuming the homogeneous pumps. We also ignore the spin degree of freedom for simplicity. The steady-state solutions we are interested in are found from:

$$P_O + P_J = g_C N_C + g_X N_X + g_{eh} N_{eh}, \tag{5}$$

$$N_X = -\frac{R}{2A} + \frac{\sqrt{(1+N_C)(R^2 + 4Ag_C N_C + R^2 N_C)}}{2A(1+N_C)}, \tag{6}$$

$$N_{Oeh} = \frac{P_O + AN_X^2(N_C + 1)}{g_{eh} + W_O}, \tag{7}$$

$$N_{Jeh} = \frac{P_J}{g_{eh} + W_J}. \tag{8}$$

First, we consider the effect of the electrical pumping only, assuming $P_O=0$. To be able to analyze the measured data of Device 2 in Fig. 1(b), we make the following changes. The electrical excitation term $P_J$ is replaced by $(JS/q)^\eta$ where, $J$ is the injected current density, $S$ is the cross-sectional area of the device active region (having dimensions of 690 nm X 40 µm), $q$ is the elementary charge and $\eta$ is a dimensionless damping factor for the electrical excitation. We see that a value of $\eta \sim 0.55$ gives the best agreement with the measured data. This parameter accounts for the effects of non-idealities in the device performance, principally nonradiative recombination, and carrier leakage from the active recombination region. The satisfactory agreement of the modeling with the measured data is achieved for the values of the parameters listed below. The relaxation rates are $\gamma_C = 1.19$ ps$^{-1}$, $\gamma_X = 0.001$ ps$^{-1}$, $\gamma_{eh} = 0.0005$ ps$^{-1}$. The relaxation rate to the condensate is $R = 3.7 \times 10^{-10}$ ps$^{-1}$, the Auger constant is $A = 3.7 \times 10^{-11}$ ps$^{-1}$. The exchange rates from the reservoirs of free carriers are $W_O = 2$ ps$^{-1}$ and $W_J = 0.01$ ps$^{-1}$. The estimations are very close to those made in [26, 31].

Next, we theoretically study optically triggered polariton lasing in Device 2, by analyzing the measured data in Fig. 1(b). To do so, we ignore Eqn. (4), we remove the term $N_{Jeh}$ from equation (2) and replace $P_O$ in Eqn. (3) by the expression: $(1- R_{NI})A_{NI}I_p/\hbar\omega$, where the symbols have the same meanings and values as previously described. Using the same set of values for the parameters as those used for the electrical excitation case, we obtain a polariton laser threshold $I_p \sim 1.6$ mW. This is in excellent agreement with the value of the incident optical power at which the photocurrent exhibits an abrupt increase in value. The polariton laser threshold, in Fig. 1(b), is defined as the value of the injection which corresponds to a ground-state occupation number of unity. To obtain this matching between the two thresholds, we have only tuned the values of two parameters to obtain $R_{NI} \sim 0.9$ and $A_{NI} \sim 0.435$. Both of these values are quite reasonable. The former corresponds to the averaged reflectance as seen by the incident wideband pump, while the latter is the overall absorbance inside the cavity. This gives an averaged effect because of absorption in the LP band and also probably some residual absorption in the free carrier reservoir.

Figures 2(a) and (b) show the measured and calculated (solid curves) variation of the output photocurrent as a function of the incident excitation. We use the same parameters as those have been used to analyze the measured forward-bias electroluminescence and the theoretically predicted photoluminescence data in Fig. 1(b). The photocurrent noticeably grows with the increase of the pump power above the threshold. We associate the growth of the photocurrent with the increase in the occupation number of the reservoir of optically excited charge carriers, $N_{Oeh}$, due to the Auger-like process. This is in agreement with our reasoning that the threshold-like behavior in the output photocurrent response is related to bosonic stimulation effects caused by the occupation of the LP ground state. This behavior should not be observed at the photon lasing threshold, as the total carrier density becomes clamped at the onset of threshold.

Finally, we theoretically consider the situation when both the optical and electrical pumps are present. Since the relaxation to the exciton (X) reservoir from the "optical" reservoir of carriers is faster than that from the "electronic" reservoir, $W_O \gg W_J$, the former pumps the X reservoir more efficiently. At the same time, it empties more quickly. Figure 3 (a) shows the dependence of $N_C$ on optical excitation $P_O$ at several values of the electrical pump power. We observe that the laser threshold remains essentially invariant of the value of the electrical pump power. The total occupation number of free carriers $N_{eh}$ calculated as a function of optical pump power for two different values of electrical pump power are shown in Fig. 3(b). The general nature of the non-linear increase in $N_{eh}$, which is a measure of the photocurrent in this dual excitation case, is similar to the measured data shown in Figs. 2(a) and (b). The same general non-linear trend is expected for zero or reverse bias. The inset of Fig. 3(b) shows the variation of $N_{Oeh}$ with optical pump power.

In conclusion, we demonstrate a new manifestation of bosonic final state stimulation in a microcavity diode in the form of a non-linear increase of the photocurrent with a distinct threshold at room temperature. The measurements have been made on a bulk GaN microcavity with lateral injection. The threshold shows excellent agreement with the measured threshold of polariton lasing in the same microcavity. The experimental results are analyzed in the framework of a model based on the coupled kinetic equations for free carriers, excitons and exciton-polaritons, in which an Auger-like process of excitonic dissociation is also included. When working on this manuscript, some of us have become aware of a similar non-linear photocurrent experiment performed on a GaAs-based microcavity device. We have applied the model also implying the stimulated Auger process as the main reason of the photocurrent non-linearity to the description of that experiment and obtained a good agreement with the data [32]. A crucial difference between the present work and the experiments reported in Ref. 32 is in the operation temperature of the polariton laser. While the data of Ref. 32 are taken at the liquid Helium temperature, here we report the results of room temperature measurements.

## Acknowledgements

The work is supported by the National Science Foundation under the Materials Research Science and Engineering Center (MRSEC) program (Grant No. DMR-1120923). The authors thank Dr. Thomas Frost for the molecular beam epitaxial growth of the device heterostructure. Device processing was carried out in the Robert H. Lurie Nanofabrication Facility. AK and ES thank Ivan Iorsh for fruitful discussions. AK acknowledges the support from the Ministry of Education and Science of The Russian Federation (project RFMEFI61617X0085). E.S. acknowledges the RFBR Grants No. 16-32-60104 and No. 17-52-10006. A.K. and E.S. acknowledge the support from the EPSRC Programme Grant on Hybrid Polaritonics No. EP/M025330/1 and the partial support from the Royal Society International Exchange Grant No. IEC/R2/170227.

**Figure Captions**

**Figure 1** (color online) (a) Schematic diagram of the microcavity polariton laser diode along with the photocurrent measurement set-up, (b) measured polariton ground-state occupation numbers recorded at zero emission angle ($k_{\parallel} \sim 0$), as a function of injected current density, for Device 2. The solid olive curve depicts the analysis of the measured data. The solid brown curve shows the theoretically calculated values for the case of optically triggered polariton lasing in the same structure.

**Figure 2** (color online) Measured photocurrent response as a function of incident optical pump power for Device 2 at a reverse bias (a) of $\sim 0.1$ V and (b) of $\sim 0.2$ V. The solid lines show the calculated values. The solid vertical arrows indicate the onset of non-linearity.

**Figure 3** (color online) (a) Polariton ground-state occupation number and, (b) total occupation number of free carriers $N_{ex} = N_{Oeh} + N_{Jeh}$ as functions of optical pump power $P_O$ at different values of electrical pump power, $P_{J1} < P_{J2} < \ldots < P_{J6}$. The inset in (b) shows the occupancy of the optically pumped reservoir of free carriers, $N_{Oeh}$ as a function of optical excitation.

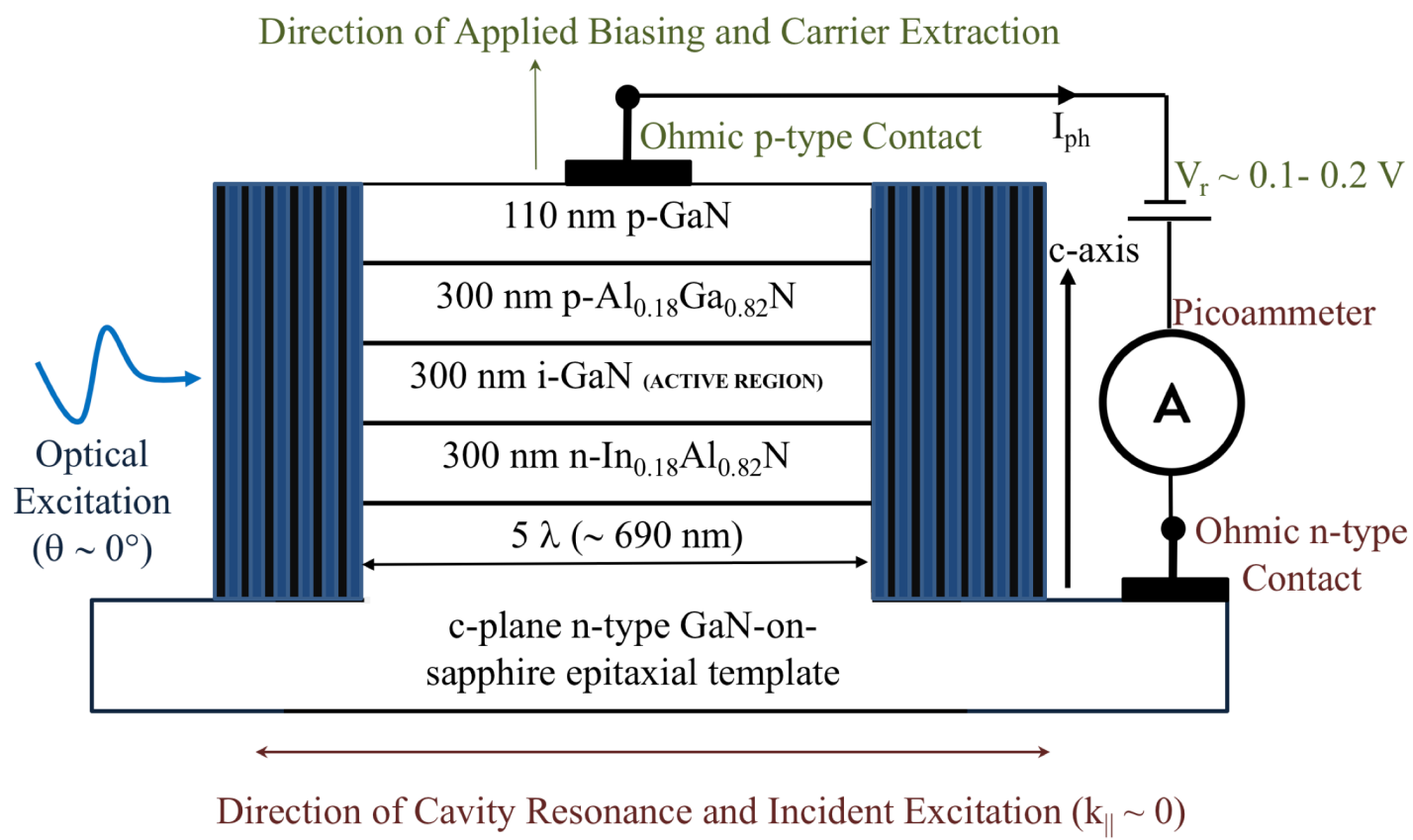

(a)

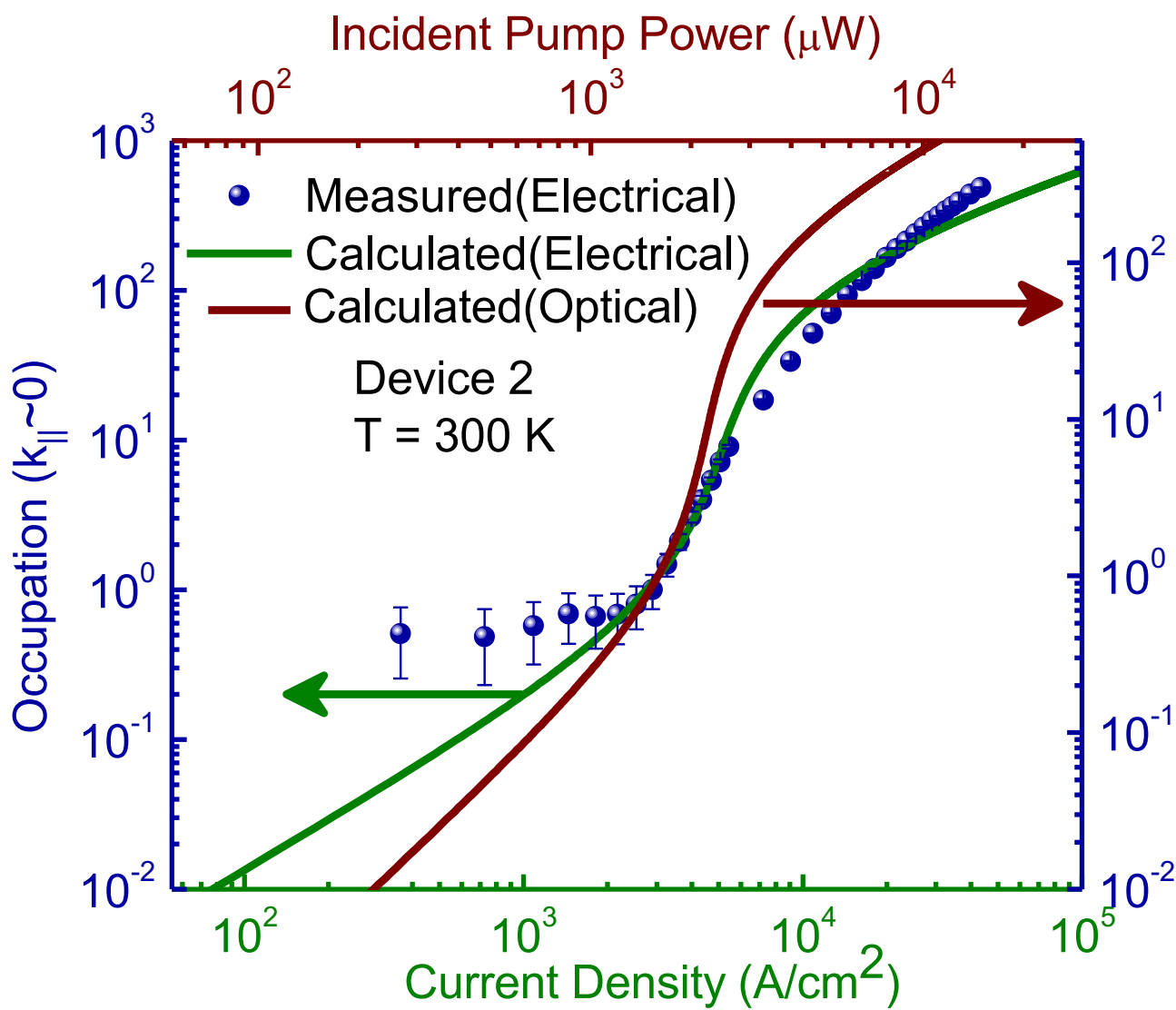

(b)

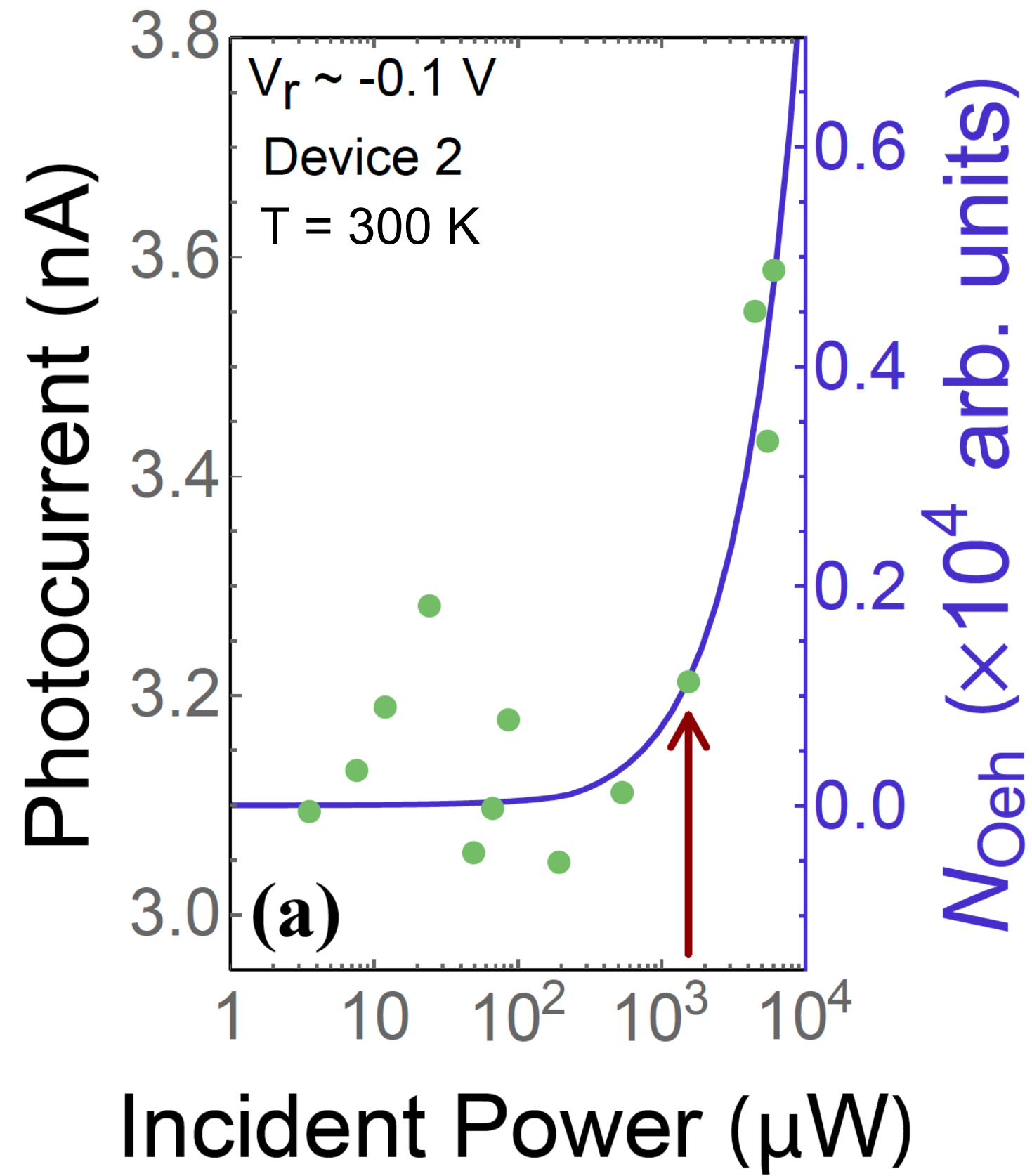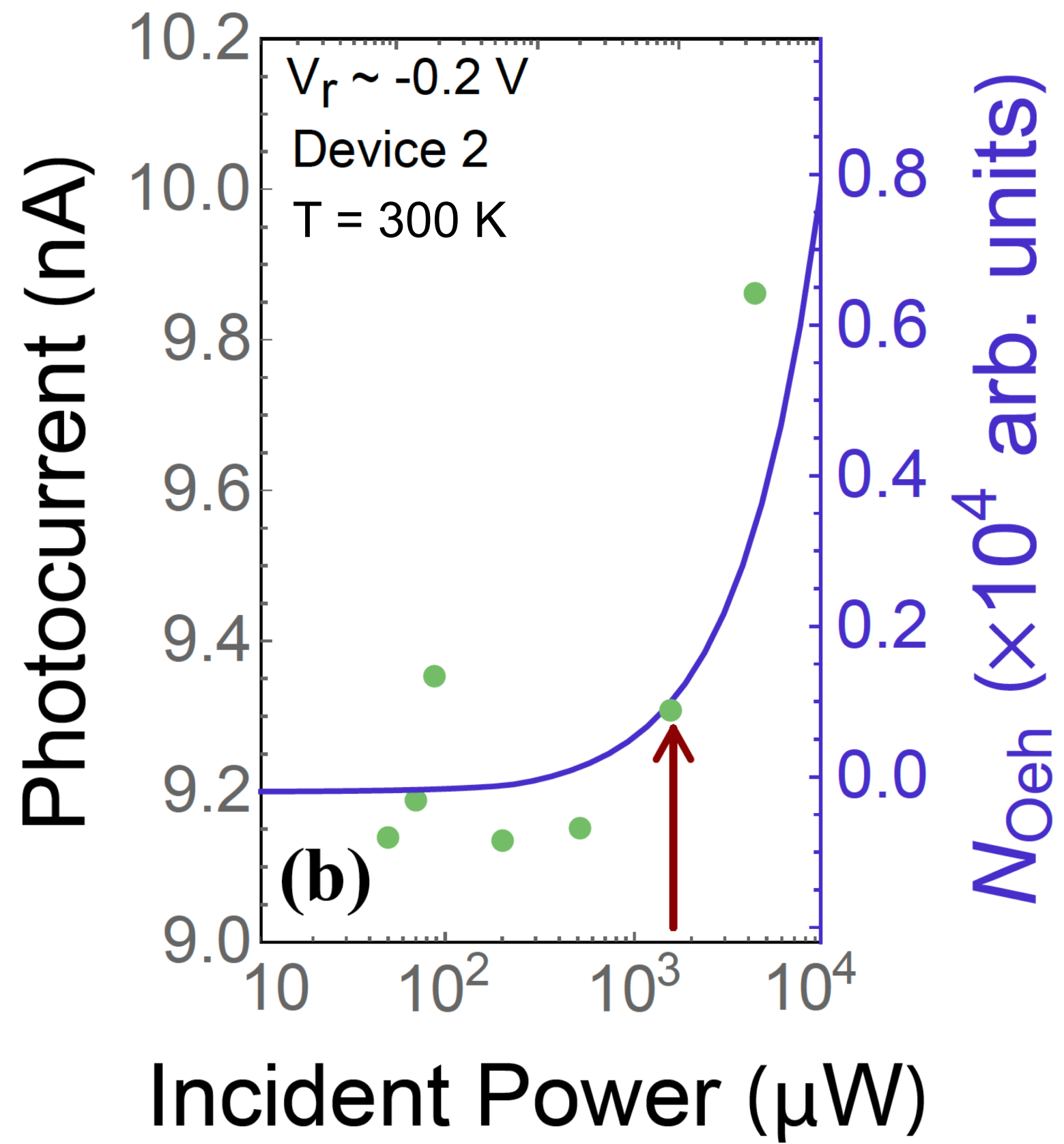

*Figure 2 of 3*

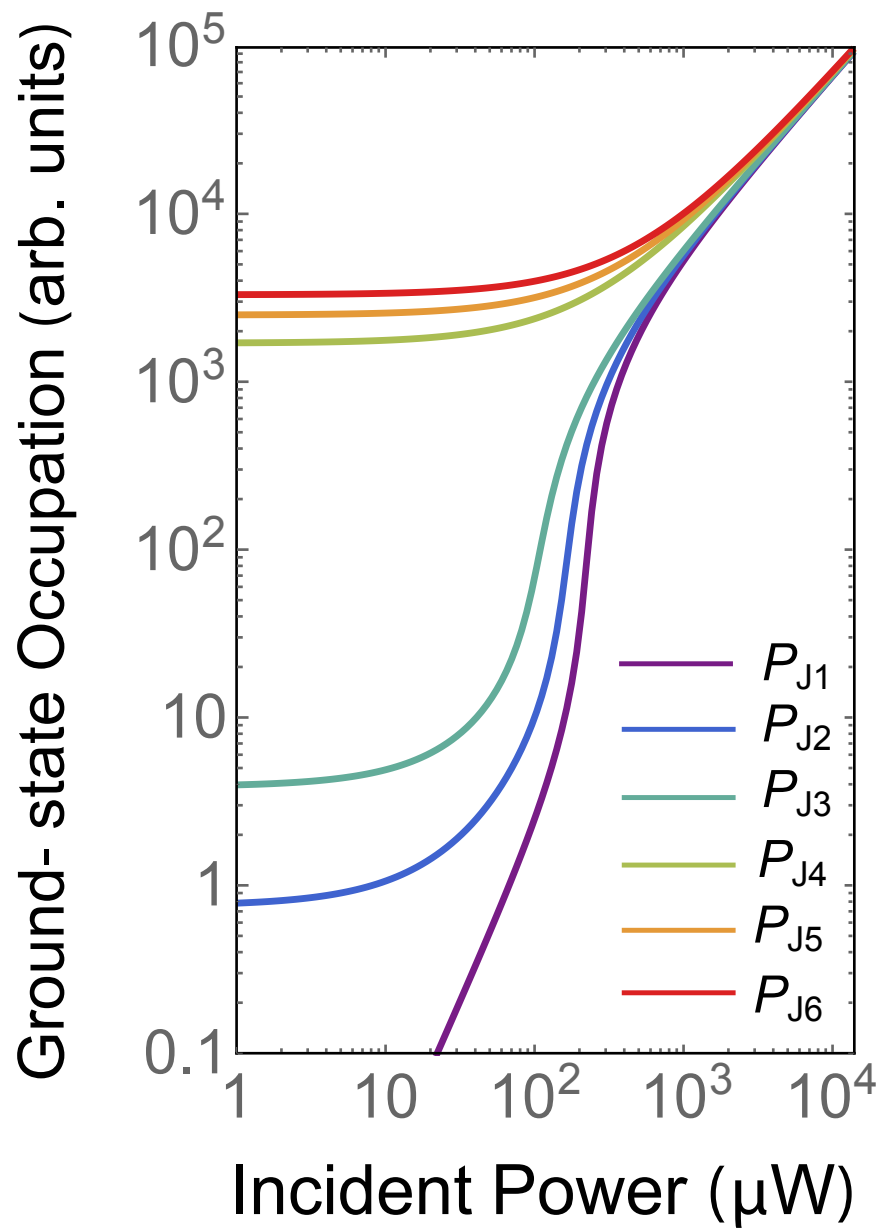 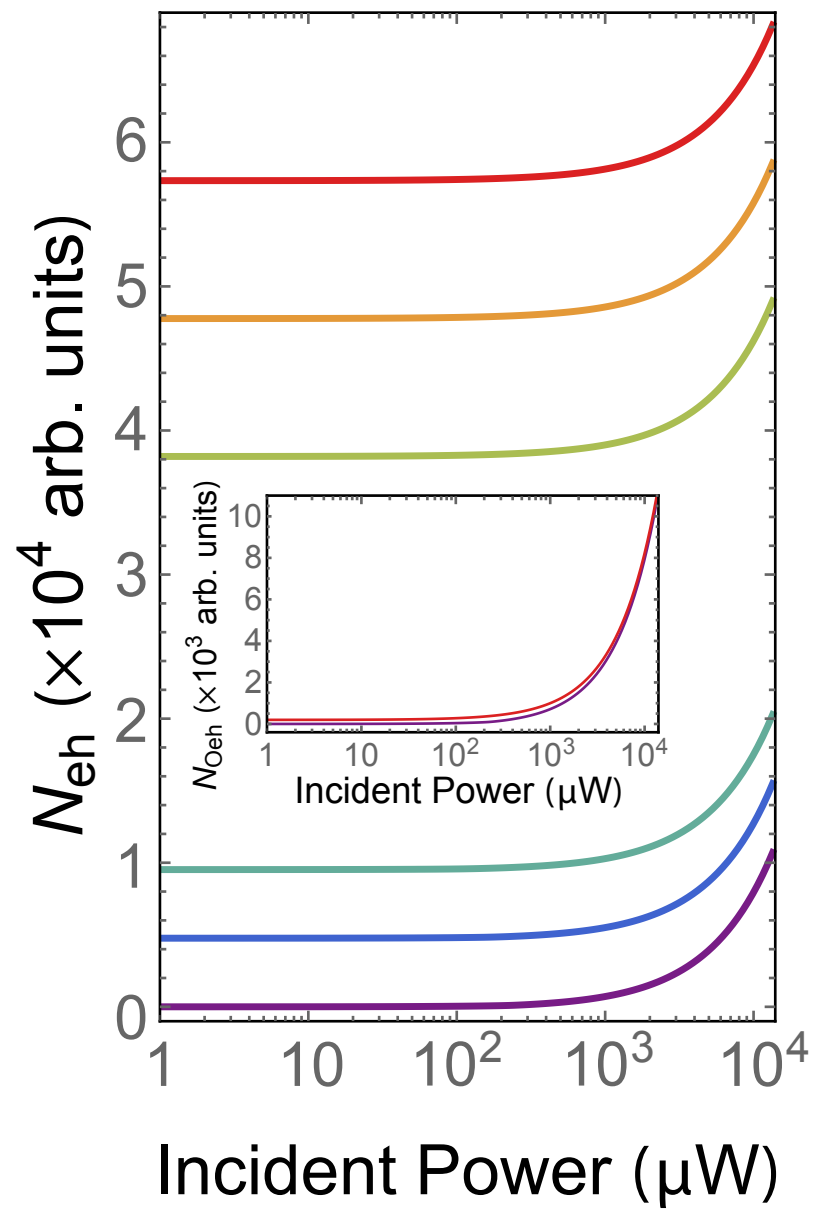

*Figure 3 of 3*

# Supplemental Material

## *1. Excitation Dependent Photocurrent Measurement Scheme*

Optical excitation-dependent photocurrent measurement are made at room temperature on the polariton diode lasers under reverse bias. Measurements have not been made under forward bias, as the diode diffusion current would overwhelm the relatively weaker photocurrent response. The diodes are illuminated at approximately near normal incidence. The excitation is provided by focusing the frequency-doubled output of a pulsed Ti: sapphire laser mode-locked at 80 MHz to a spot size of ~ 10 µm in diameter (by using an UV objective lens) upon one of the device facets. The diode reverse current is recorded with a high-precision digital pico-ammeter having a resolution of ~ 100 fA. The frequency of the excitation laser source was spectrally tuned to coincide with the LP mode energy of ~ 3.378 eV (~ 367 nm) observed at the threshold. This has been done to avoid directly exciting the electron-hole plasma reservoir, which could potentially mask the non-linearity in the photocurrent response from the stimulated excitonic dissociation process, as subsequently discussed. Nevertheless, we do believe that the strong laser background does also bring about true band-to-band excitations in the fermionic reservoir. The blue-shift of the LP mode is not taken into account to spectrally tune the excitation wavelength as a function of incident optical pumping intensity. However, since the spectral linewidth (full-width at half maximum) of the excitation laser is ~ 2.1 nm (or equivalently ~ 20 meV), appreciable excitation of the ground-state lower-polariton (LP) states, which are slightly red-shifted below threshold and very slightly blue-shifted above threshold, takes place. The rather broad linewidth of the excitation is a consequence of its pulsed nature. This particular configuration, i.e., of using normal incidence pulsed excitation has been motivated by our desire to study the device in the context of a more realistic practical situation;

it is more convenient to excite a diode at normal incidence to its facets as opposed to any other particular fixed angle of incidence. Further, the combination of the ~ 9 meV (half-width at half maximum) laser tail, along with the angular divergence of the incident beam, should also excite some of the higher energy lower-polariton states. It is certainly probable that stimulated scattering from higher energy states along the lower-polariton dispersion to the ground state should be possible in the present configuration, assuming that the ground state occupation number is greater than unity, above a certain excitation value.

## 2. *Light-Current and Optical Excitation Dependent Photocurrent Characteristics of Device 1*

The light (output)-current (L-I) characteristics of Device 1, at zero angle of emission ($k_\parallel \sim 0$), are determined by recording the output lower polariton (LP) electroluminescence intensities by a photomultiplier tube after spectral filtering through an imaging monochromator, as a function of continuous wave injection current density are shown in Fig. S1. The polariton lasing threshold current density is ~ 1993 A/cm$^2$. The sub-threshold slope of these characteristics, when plotted in a double logarithmic scale as in Fig. S1, is ~ 0.45 whereas, the non-linear slope above the threshold is ~ 2.95.

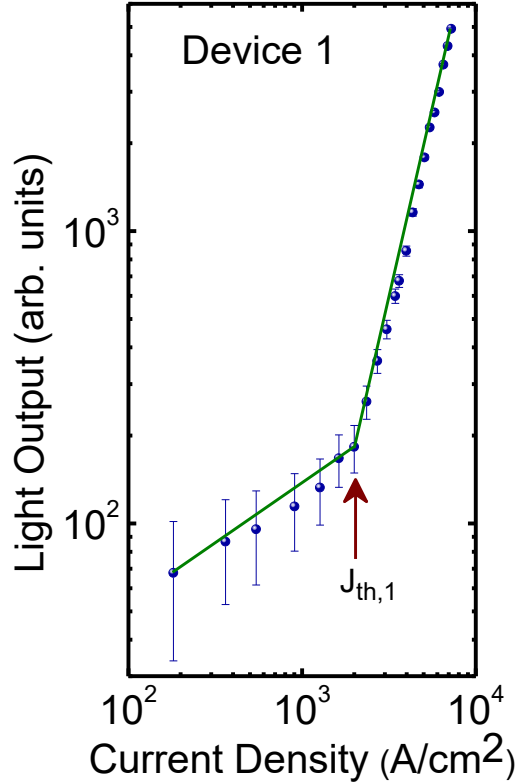

**Figure S1.** LP electroluminescence intensities recorded at zero emission angle ($k_{\parallel} \sim 0$), as a function of injected current density. The solid lines are guides to the eye. The solid vertical arrow indicates the onset of non-linearity.

Excitation-dependent photocurrent measurements were also performed on this device, in the same manner as has been described in Section 1. The data are shown in Fig. S2. The onset of the non-linearity in the photocurrent response is at an incident optical power $I_p \sim 1$ mW. The non-linearity in the output photocurrent response is relatively weaker as compared to that observed in Device 2. This is probably due to the fact that the non-linearity in this particular device is inherently weaker, as compared to that in Device 2, and this has been discussed in the main text.

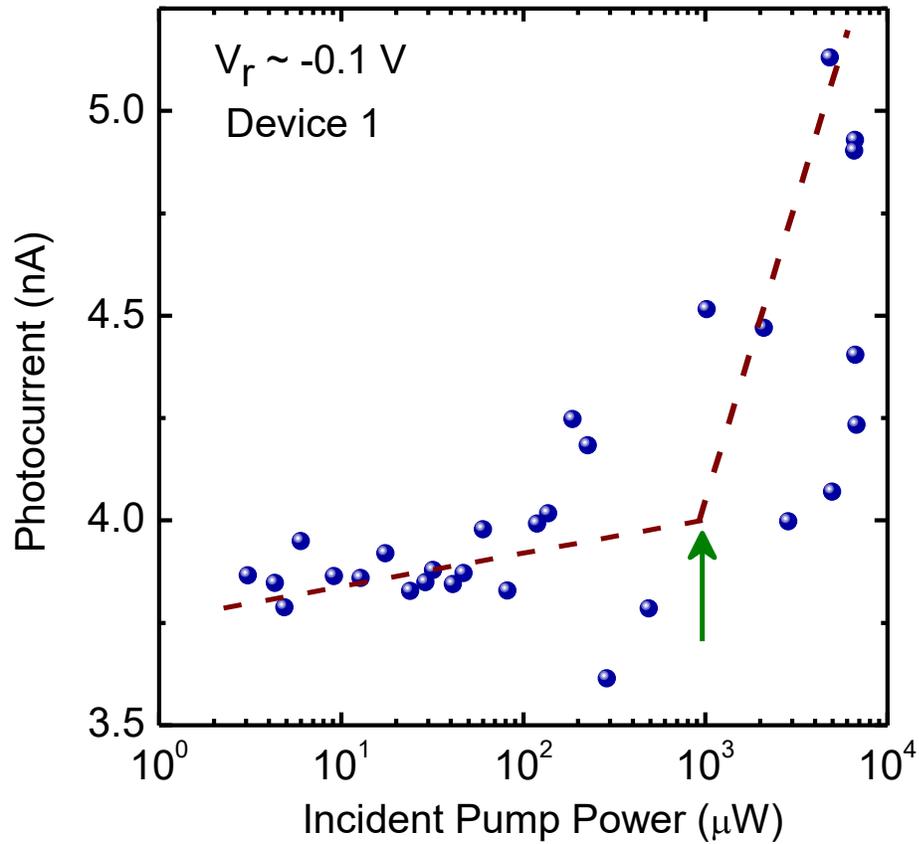

**Figure S2.** Measured photocurrent response as a function of incident optical pump power for Device 1 at a reverse bias of ~ 0.1 V. The dashed lines are guides to the eye. The solid vertical arrow indicates the onset of the non-linear photocurrent response at ~ 1 mW.

## 3. Light-Current Characteristics of a Representative Device Prior to Degradation

Figure S3 shows the light (output)-current characteristics of a representative polariton laser device. This device has been extensively characterized in one of our previous publications [SR1]. These characteristics were recorded immediately after the conclusion of all processing steps. The excitation-dependent output photocurrent measurements, detailed in the main text as well as in the Supplemental Material, were made on similar structurally identical devices all originating from the same epitaxial sample. Further, all the devices were processed simultaneously. The polariton lasing threshold current density is $J_{th,1} \sim 262.5$ A/cm$^2$. The nonlinear region of the electroluminescence is characterized by a slope of ~13.7, and the enhancement of the output coherent luminescence over the active lasing regime is ~ 3.5 orders of magnitude. The solid blue curve shows the calculated dependence of the polariton ground-state occupation number and the agreement of the measured data with the calculated data is reasonably good.

Conventional semiconductor photon lasing was also observed in the same device at an injected current density of $J_{th,2} \sim 36.8$ kA/cm$^2$. This has been shown in the inset of Fig. S3. The observation of two distinct thresholds, ~ 2 orders of magnitude apart in injected carrier density, is indicative of the very different natures of the origin of the nonlinear emissions. It has been our observation that photon lasing, in these double heterostructure GaN-based diodes, is only achieved in the very best devices, wherein the degree of non-linearity of the LP emission characteristics is reasonably high. This probably implies a lower density of defects and other non-radiative recombination and trapping centers, which translates to a lower value for the intra-cavity loss.

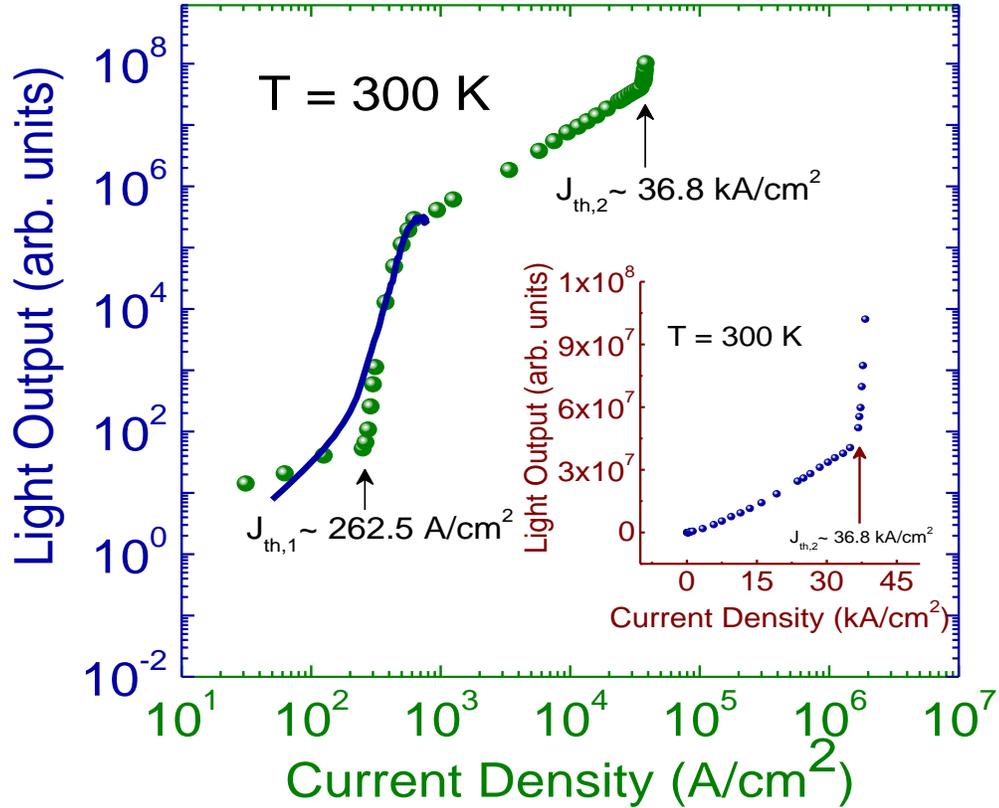

**Figure S3.** LP electroluminescence intensities recorded at zero emission angle ($k_{\parallel} \sim 0$) as a function of injected current density. The solid blue line represents theoretically calculated values. The solid vertical arrows indicate the onset of the two distinct non-linearities. The inset shows the same data, plotted in a double linear scale, to highlight the photon lasing regime. The experimental data, shown in this figure, have been reported by us in a previous publication [SR1].

## 4. Angle-resolved Electroluminescence Measurements and Strong Coupling Characteristics

The strong coupling characteristics of a GaN-based device, similar to those which have discussed in the main text and the Supplemental Material, are discussed here for completeness. This particular device has also been fully characterized in one of our previous publications [RS2]. The angle-resolved electroluminescence data, which are discussed in the present section, have been previously reported in the Supplementary Materials of Refs. [SR2] and [SR3]. A short description of the angle-resolved measurements is as follows. To ascertain the nature of the exciton-photon strong-coupling characteristics, of the GaN-based devices, angle-resolved electroluminescence measurements were made at a continuous-wave (CW) injection close to the polariton laser threshold (~ 0.95 $J_{th}$) using a digital readout angular mount having an angular precision of ~ 0.1°. The spectra were recorded employing a combination of a 0.75 m Czerny-Turner imaging monochromator (with a spectral resolution of ~ 0.03 nm at ~ 435.8 nm) and a photomultiplier tube for a range of different emission angles. LP resonant peaks are observed below the exciton energy at all angles and the LP peaks tend to approach the exciton energy at higher angles (Fig. S4 (a)). The corresponding polariton dispersion characteristics, which are calculated using a 2 X 2 coupled harmonic oscillator model, are shown in Fig. S4 (b), from which we deduce the cavity-to-exciton detuning δ and Rabi splitting Ω to be ~ – 13 meV and ~ 33.9 meV, respectively.

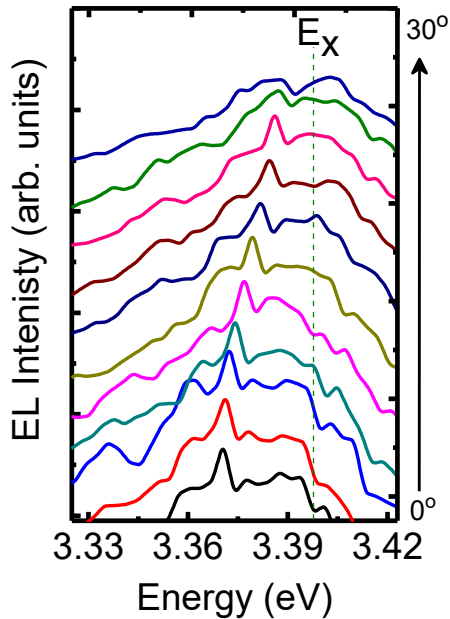

(a)

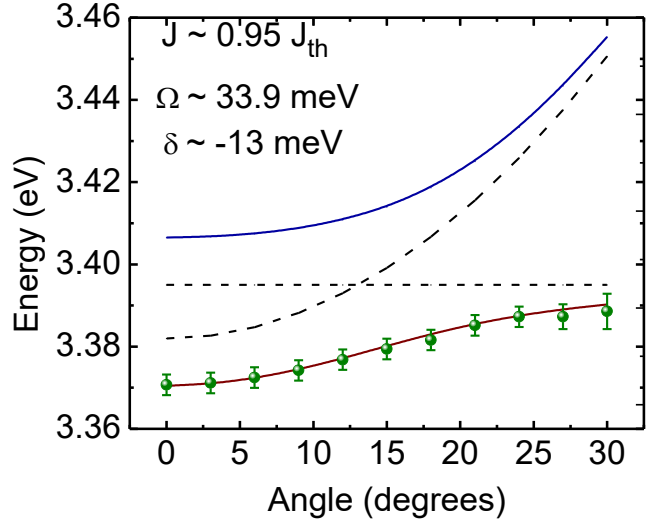

(b)

**Figure S4.** (a) Measured angle-resolved electroluminescence spectra of an identical GaN-based device. The spectra have been vertically offset for clarity. The dashed line represents the excitonic transition, (b) corresponding polariton dispersion characteristics. This figure has been reproduced from the Supplementary Materials of Refs. [SR2] and [SR3].

## 5. *On the Validity of the Quasianalytical Model Based on the Coupled Rate Equations for an Exciton-Polariton Device*

The rate equations (1)–(4) in the main text allow one to qualitatively understand the kinetics of particles in the considered exciton-polariton device. The model accounts for the effects of the polariton condensation threshold as well as the mechanisms of feeding the polariton state due to the phonon scattering into it and due to the Auger-like process in the exciton reservoir. An alternative model allowing characterization of the exciton-polariton device above the threshold is based on the equation for the order parameter $\Upsilon$ which is the many-body wave function of the polariton condensate, coupled to the equations for the occupancies of the exciton and electron-hole reservoirs:

$$i\frac{dY}{dt} = \left[a_C |Y|^2 + a_X N_X + \frac{i}{2}(RN_X + AN_X^2 - g_C)\right]Y, \tag{S1}$$

$$\frac{dN_X}{dt} = W_O N_{Oeh} + W_J N_{Jeh} - [g_X + R|Y|^2 + 2A|Y|^2 N_X]N_X, \tag{S2}$$

$$\frac{dN_{Oeh}}{dt} = P_O + A|Y|^2 N_X^2 - (g_{eh} + W_O)N_{Oeh}, \tag{S3}$$

$$\frac{dN_{Jeh}}{dt} = P_J - (g_{eh} + W_J)N_{Jeh}. \tag{S4}$$

In Eqn. (1) $\alpha_C$ and $\alpha_R$ are the nonlinear constants describing the polariton-polariton and polariton-exciton interactions. The other parameters are the same as in Eqs. (1)–(4) in the main manuscript.

Figure S5 shows the polariton occupation number $N_C$ in the steady state as a function of the electrical pump power $P_J$, assuming the above-mentioned values. The dependence is typical for polariton lasers. The dependence above the threshold calculated from Eqs. (S1)–(S4) (blue dots) perfectly matches one calculated from Eqs. (1)–(4) (solid curve).

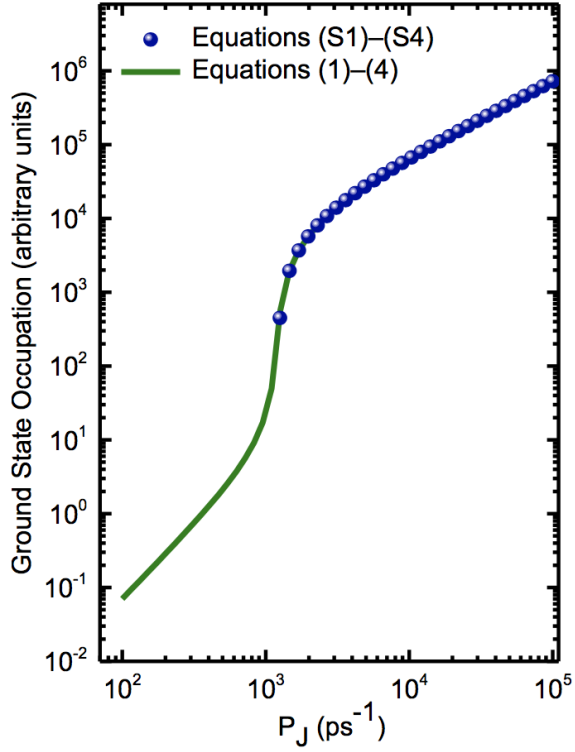

**Figure S5**. Calculated polariton ground state occupation numbers as a function of the electrical pump power $P_J$. The green line shows the steady-state occupation number $N_C$ calculated from Eqs. (1) through (4). The blue dots show the dependence above the threshold calculated from Eqs. (S1)–(S4). The interaction constants are taken $\alpha_C = 0.0001 \text{ps}^{-1}$ and $\alpha_X = 2\alpha_C$. The other parameters are the same as in the man text.

## 6. *Spectral Characteristics of the Excitation Source*

As discussed in Section 1 of the Supplemental Material, the light source used to excite the exciton-polariton device is the relatively broadband frequency-doubled output of a Ti: Sapphire laser. The spectral output of the excitation source is shown in Fig. S6. The full-width-at half-maximum of the spectrum is ~ 2.1 nm or equivalently ~ 19.7 meV. The peak wavelength of the excitation source is ~ 367.02 nm. The laser source was tuned to this wavelength to match the peak electroluminescence emission wavelength of the polariton laser diode on which the photocurrent measurements have been made. The emission spectrum of the polariton laser diode was recorded just at the onset of the polariton lasing threshold ($J \sim J_{th}$).

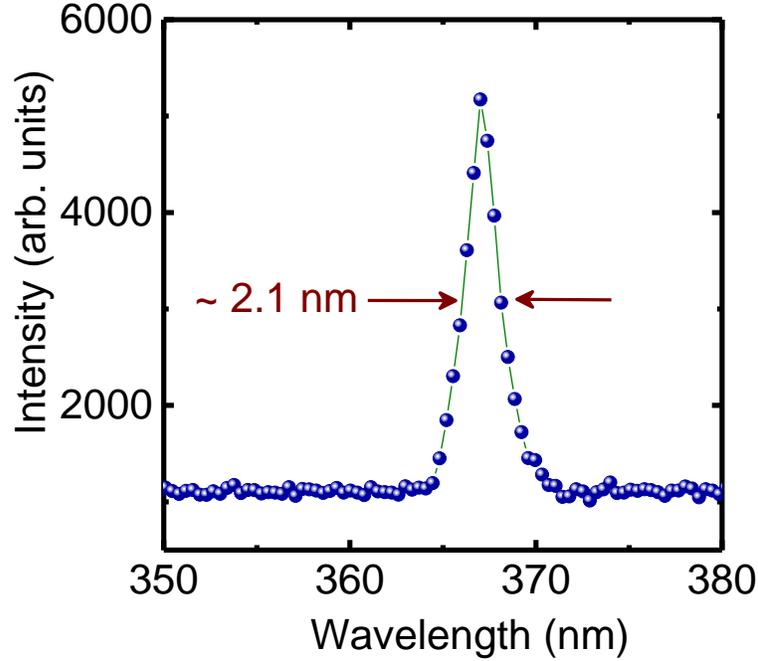

**Figure S6.** Spectrum of the doubled output of the Ti: Sapphire laser used to excite the microcavity diode at near normal incidence.

We assign the peak having the maximum signal amplitude as that originating from polariton emission. The data are shown in Fig. S7. The linewidth of the overall main emission is ~ 1 meV and is centered at ~ 3.378 eV (~ 367 nm). As has been previously reported in the literature, GaN-based polariton lasers often exhibit multiple transverse modes [SR4], which probably explains the 2 relatively closed spaced resonances. A simplified analysis of the data in Fig. S7 using multi-peak Gaussian modeling is shown in Fig. S8. The error bars in Fig. S8 indicate the variation of the measured noise floor from the baseline obtained from the analysis, and thus is indicative of the noise intrinsic to the measurement. The spectrum was acquired with a 0.75 m Czerny-Turner type imaging monochromator with ~ 0.023 nm spectral resolution at ~ 435.835 nm. Both the entrance and exit slit widths were reduced to the minimum value of 30 μm.

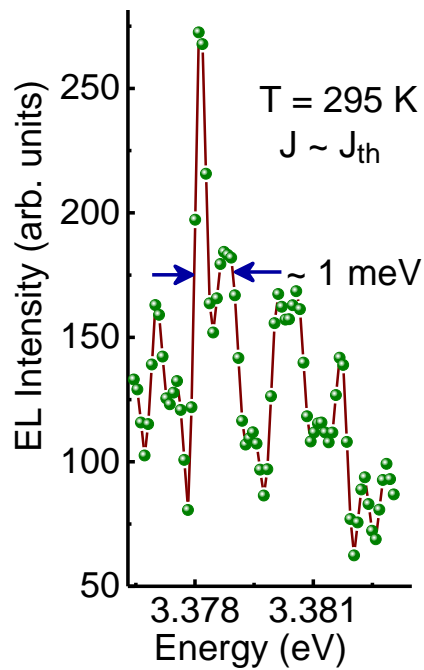

**Figure S7.** Electroluminescence spectrum recorded, at zero angle of emission, at the onset of non-linearity (J ~ J$_{th}$) in Device 2. The solid brown curve is a guide to the eye.

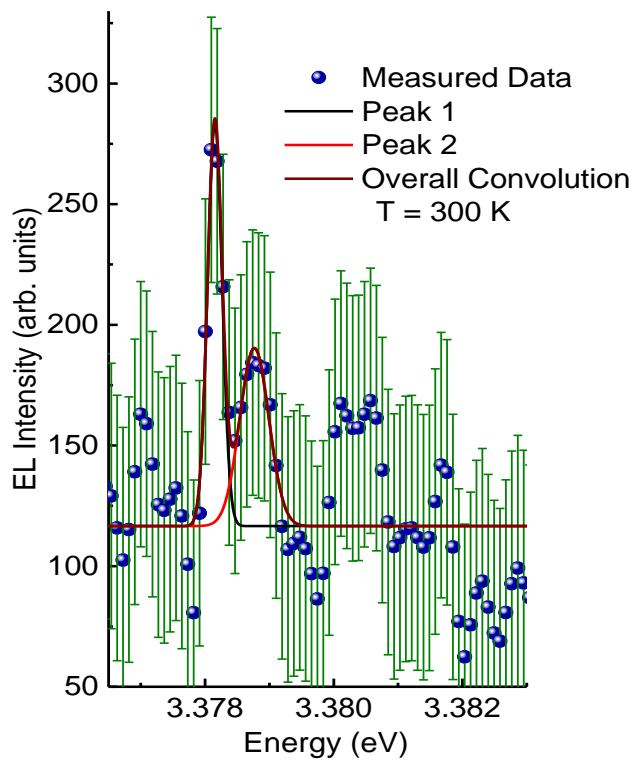

**Figure S8.** Analysis of the data in Fig. S7. The two constituent resonances have linewidths of ~ 300 µeV and ~ 500 µeV. The resolution of the measurement is ~ 150 µeV.

**References for the Supplementary Material:**